\documentclass[onecolumn]{revtex4-1}

\usepackage[version=3]{mhchem} 
\usepackage{graphicx}
\usepackage{amssymb}
\usepackage{epstopdf}
\usepackage{epsfig}
\usepackage{color}
\usepackage{amsmath}
\usepackage{amsfonts}
\usepackage{graphicx}
\usepackage{fancybox}
\usepackage{subfigure}
\usepackage{notes2bib}
\usepackage{float}
\usepackage{placeins}

\newcommand{\pvn}{P_v(N)}

\newcommand{\gcav}{G_{\rm cav}}
\newcommand{\ghyd}{G_{\rm hyd}}
\newcommand{\gtr}{G_{\rm trans}}

\newcommand{\scav}{S_{\rm cav}}

\newcommand{\fig}[2]{\textbf{Figure~\ref{#1}{#2}}}

\begin{document}

\title{Understanding Hydrophobic Effects: Insights from Water Density Fluctuations}

\author{Nicholas B. Rego}
\affiliation{Department of Chemical and Biomolecular Engineering, University of Pennsylvania, Philadelphia, PA 19104, USA}

\author{Amish J. Patel}\email[]{amish.patel@seas.upenn.edu}
\affiliation{Department of Chemical and Biomolecular Engineering, University of Pennsylvania, Philadelphia, PA 19104, USA}

\begin{abstract}
The aversion of hydrophobic solutes for water drives diverse interactions and assemblies across materials science, biology and beyond. 
Here, we review the theoretical, computational and experimental developments which 
underpin a contemporary understanding of hydrophobic effects.
We discuss how an understanding of density fluctuations in bulk water can shed light on 
the fundamental differences in the hydration of molecular and macroscopic solutes; 
these differences, in turn, explain why hydrophobic interactions become stronger upon increasing temperature.
We also illustrate the sensitive dependence of surface hydrophobicity on the chemical and topographical patterns the surface displays, 
which makes the use approximate approaches for estimating hydrophobicity particularly challenging.
Importantly, the hydrophobicity of complex surfaces, such as those of proteins, which display nanoscale heterogeneity, can nevertheless be characterized using interfacial water density fluctuations; 
such a characterization also informs protein regions that mediate their interactions.
Finally, we build upon an understanding of hydrophobic hydration and the ability to characterize hydrophobicity to inform the context-dependent thermodynamic forces that drive hydrophobic interactions and the desolvation barriers that impede them. 
\end{abstract}

\maketitle


\section{INTRODUCTION}
Hydrophobic effects, which refer to the aversion of non-polar solutes for water, or the affinity of such solutes for one another in aqueous solutions, play an important role in diverse disciplines, ranging from interfacial science and supramolecular chemistry, to soft matter and biophysics~\cite{Ball:CR:2008,Jamadagni:ARCBE:2011,Gibb:ARPC:2016,Ben-Amotz:ARPC:2016,Berne:ARPC:2009,Dill:ARB:2005}.
For example, amphiphilic molecules or particles, which possess both non-polar and polar regions, 
can ameliorate unfavorable interactions between oil and water, thereby giving rise to detergency and stabilizing emulsions~\cite{Tanford:book,Haase:ACSNano:2016}.
Moreover, given that half the naturally occurring amino acids are non-polar, 
the hydrophobic effect also drives numerous biomolecular interactions and assemblies~\cite{Dobson:Nature:2003,Levy:ARBBS:2006,Krone:JACS:2008,Thirumalai:ARB:2010}. 
The central challenge in understanding hydrophobic hydration, characterizing hydrophobicity, and anticipating hydrophobic interactions lies in capturing how hydrophobic solutes perturb water structure in their vicinity.
Being a highly cohesive liquid, water tries to minimize the disruption of its structure 
by collectively reorganizing its hydrogen bonding network~\cite{Chandler:lecture}; 
however, the extent to which it succeeds can depend in non-trivial ways on solute properties, 
such as its size, shape, curvature, and the arrangement of its chemical groups.

In this review, we discuss the fundamental insights which underpin a contemporary understanding 
of hydrophobic effects. 
Foremost amongst them is the dependence of hydrophobic hydration on solute size.
In particular, because water can hydrogen bond around sub-nm hydrophobic solutes, but not around larger solutes, molecular and macroscopic solutes perturb water structure differently~\cite{Chandler:Nature:2005}. 
An understanding of density fluctuations in bulk water can shed light on these differences, 
and explain how the hydration free energy of a small solute is dominated by the entropic cost of 
constraining the hydrogen bond network of its hydration waters, and scales as solute volume;
whereas, the hydration free energy of a large solute is dominated by the enthalpic cost of 
broken hydrogen bonds in its hydration shell, and scales as its surface area~\cite{Chandler:lecture}. 
Similarly, an understanding of water density fluctuations in the vicinity of a surface 
serves to characterize its hydrophobicity.
In particular, the free energetic cost of displacing interfacial waters to create a cavity next to a surface provides a general, unambiguous measure of its hydrophobicity; the easier it is to create an interfacial cavity, the more hydrophobic the surface~\cite{Godawat:PNAS:2009,Patel:JPCB:2010,Patel:PNAS:2011,Patel:JPCB:2012,Patel:JPCB:2014}.
Such a measure can also be used to estimate the hydrophobicity of complex surfaces, such as those of proteins, which display chemical and topographical heterogeneity at the nanoscale.
The use of interfacial fluctuations to characterize the hydrophobicity of complex surfaces has highlighted that
hydrophobicity can be remarkably sensitive to curvature and chemical patterning~\cite{Giovambattista:PNAS:2008,Giovambattista:PNAS:2009,Acharya:Faraday:2010,Mittal:Faraday:2010,Xi:PNAS:2017}, and that approximate approaches for estimating hydrophobicity, based on surface area or additivity, are doomed to fail for all but the simplest of systems~\cite{Fennell:JSP:2011,Harris:PNAS:2014,Wang:PNAS:2011,Factorovich:JACS:2015}.
We then discuss how an understanding hydrophobic hydration and the ability to characterize hydrophobicity inform the thermodynamic forces that drive hydrophobic interactions and assemblies~\cite{Netz:PNAS:2015,Rego:PNAS:2021}.
We also discuss how water density fluctuations in confinement between hydrophobic solutes can influence the kinetics and pathways of hydrophobic assembly~\cite{Miller:PNAS:2007,Setny:PNAS:2013,Mondal:PNAS:2013,Tiwary:PNAS:2015,Jiang:JPCB:2019,Dhabal:JPCB:2021}.

Throughout, we draw on insights obtained from theoretical and computational studies, and make connections to experiments.
As with any vibrant field, our understanding of hydrophobic effects has been shaped by numerous dedicated researchers and countless insightful studies.
Regrettably, we are unable to discuss many important findings here, including several that have inspired our work, and shaped our outlook on the subject.
Fortunately, we are able to point the interested reader to a number of recent reviews, 
which focus on different aspects of hydrophobic effects, and ought to serve as excellent 
complements to our discussion here~\cite{Berne:ARPC:2009,Jamadagni:ARCBE:2011,Giovambattista:ARPC:2012,Baron:ARPC:2013,Ben-Amotz:ARPC:2016,Gibb:ARPC:2016,Bellissent-Funel:CR:2016,Monroe:ARCBE:2020}.
%

\section{HYDRATION OF SMALL AND LARGE HYDROPHOBIC SOLUTES}
When a hydrophobic solute is hydrated, i.e., transferred to water from vapor or oil (\fig{fig:hyd}{a}), 
it disrupts the inherent structure of water~\cite{Hummer:JPCB:1998,Southall:JPCB:2002}.
Water molecules in the vicinity of the solute try to minimize 
this disruption by collectively reorganizing their hydrogen bonding network; 
however, the extent to which they succeed depends in interesting and non-trivial ways 
on the size and shape of the hydrophobic solute.
It is this collective reorganization of hydration waters, and its complex dependence on solute properties, 
which confers hydrophobic effects with their many exotic properties. 

\vspace{-0.05in}
\subsection{How Water Responds to Molecular and Macroscopic Hydrophobic Solutes}
\begin{figure}[h]
\includegraphics[width=1.\textwidth]{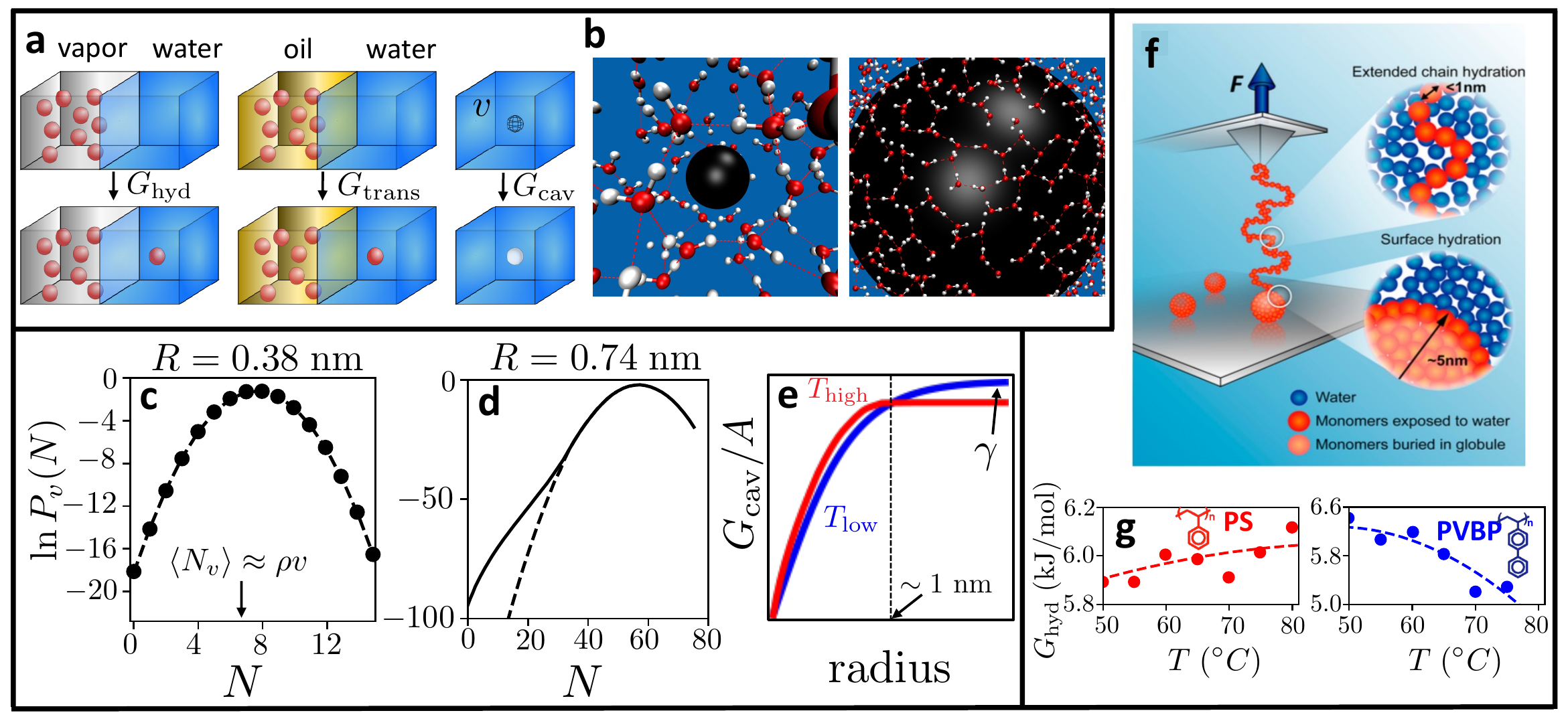}
\vspace{-0.05in}
\caption{
Hydrophobic hydration, density fluctuations in bulk water, and solute size-dependent crossover.
(a) The hydration (i.e., dissolution in water from vapor or oil) of a non-polar solute (red sphere) disrupts water structure in ways that are remarkably similar to the creation of a cavity by emptying a solute-sized volume, $v$, in bulk water.
(b) The hydrogen bond network of water is strained, but remains intact near small solutes, such as methane (left); 
in contrast, water is unable to form four hydrogen bonds near extended hydrophobic solutes (right).
(c) The probability, $\pvn$, of observing $N$ waters in small observation volumes, $v$, which contain only a few waters on average, is Gaussian (dashed line) for all $N$~\cite{Chandler:lecture}. 
(d) For larger probe volumes, although $\pvn$ is Gaussian near the mean, 
low-$N$ fluctuations are enhanced~\cite{Chandler:lecture}.  
(e) Schematic illustrating that the hydration free energy, $\ghyd$, 
of small spherical solutes scales with their volume, 
whereas for large solutes, $\ghyd$ is determined by interfacial physics, and scales with solute area, $A$;
the crossover between the two regimes occurs at a solute size of roughly $1$~nm~\cite{Chandler:lecture}. 
Moreover, small solute hydration is entropically unfavorable, so $\ghyd$ increases as temperature is increased from $T_{\rm low}$ (black) to $T_{\rm high}$ (red); in contrast, for large solutes, $\ghyd$ decreases as temperature is increased. 
(f) By using an AFM tip to pull on polymers, Li and Walker~\cite{Li:PNAS:2011} forcibly hydrated hydrophobic monomers; the requisite force informed the monomer hydration free energy. 
(g) Hydration free energies, $\ghyd$, are shown as a function of temperature for two monomer side-chains. 
For the smaller monomer, $\ghyd$ increases with $T$ (left), whereas for the larger monomer, 
it decreases above $50^\circ$C (right)~\cite{Li:PNAS:2011}.
Panels c-e adapted with permission from Reference~\citenum{Chandler:lecture}. 
Panels f and g adapted with permission from Reference~\citenum{Li:PNAS:2011}; copyright 2011 National Academy of Sciences.
}
\vspace{-0.1in}
\label{fig:hyd}
\end{figure}

The manner in which molecular and macroscopic hydrophobic solutes
perturb water structure is fundamentally different (\fig{fig:hyd}{b}).
Water molecules near a small non-polar solute, such as methane, are able to maintain all their hydrogen bonds.
However, accommodating the solute severely constrains the hydrogen bonding network of its hydration waters, 
and results in a large negative entropy of hydration~\cite{Chandler:Nature:2005}.
In contrast, the hydration waters of a macroscopic hydrophobic solute, such as a colloidal particle, 
are simply incapable of participating in four hydrogen bonds like those in bulk water;
instead, they sacrifice a hydrogen bond like water molecules at a water-vapor interface~\cite{Lee:JCP:1984,Willard:JPCB:2010}. 
%
%
Given these fundamental differences in the hydration of molecular and macroscopic solutes, 
how do we characterize the extent to which diverse hydrophobic solutes perturb water structure in their vicinity? 
The answer to this question lies in an understanding of density fluctuations in bulk water;
fluctuations provide a quantitative framework for characterizing the disruption of water structure by hydrophobic solutes, and for informing their hydration free energies~\cite{Hummer:PNAS:1996,Garde:PRL:1996,Patel:JSP:2011}.
%

%
%

\subsection{Water Density Fluctuations Inform Hydrophobic Hydration}
Non-polar solutes exclude water molecules from the regions they occupy, 
and attract their hydration waters through dispersive interactions; 
however, the latter are relatively weak, and do not perturb water structure substantially~\cite{Remsing:JCP:2015}.
Consequently, hard solutes or cavities, which simply exclude water molecules (\fig{fig:hyd}{a}), 
have been extensively studied as idealized hydrophobic solutes~\cite{Weeks:ARPC:2002,Ashbaugh:RMP:2006,Remsing:JPCB:2013}.
Importantly, because hydrating a hard solute or creating a cavity 
is equivalent to emptying a solute-sized volume $v$, 
the hydration free energy of idealized solutes, $\gcav =  - k_{\rm B} T \ln P_v(N\to0)$, 
where $k_{\rm B}$ is Boltzmann's constant, $T$ is temperature, 
and $\pvn$ is the probability of observing $N$ waters in $v$~\cite{Widom:JCP:1963,Beck:book}. 

Using molecular simulations, Hummer et al.~\cite{Hummer:PNAS:1996} showed that the statistics of water number fluctuations, $\pvn$, in small observation volumes are Gaussian for all $N$, i.e., 
\begin{equation}
\pvn \approx  \frac{1}{\sqrt{2\pi\sigma^2}} \cdot \exp \bigg[- \frac{(N-\langle N_v \rangle)^2}{2\sigma^2} \bigg],  
\label{eq:G_small}
\end{equation}
where $\langle N_v \rangle = \rho v$ and $\sigma^2 \equiv \langle \delta N_v^2 \rangle$ are the mean and variance of $\pvn$, respectively, and $\rho$ is the density of bulk water (\fig{fig:hyd}{c}).
Along with famed theoretical developments, such as the Percus-Yevick theory of hard spheres and the Pratt-Chandler theory of hydrophobic hydration, the above finding is underpinned by the fact that in a dense fluid, small fluctuations in density obey Gaussian statistics~\cite{Chandler:PRE:1993,Pratt:JCP:1977,Pratt:ARPC:2002}.
Importantly, it led to a relationship between $\gcav$ and the moments of $\pvn$~\cite{Hummer:PNAS:1996,Garde:PRL:1996}:
\begin{equation}
\gcav \approx  k_{\rm B} T \bigg[ \frac{\rho^2 v^2}{2 \sigma^2} + \frac{1}{2}\ln(2\pi\sigma^2) \bigg] \approx \frac{k_{\rm B} T}{2} \cdot \frac{\rho^2 v^2}{\sigma^2}.
\label{eq:G_small}
\end{equation}
To a reasonable approximation, $\sigma^2 = \langle \delta N_v^2 \rangle \propto \langle N_v \rangle \propto v$.
Thus, $\gcav \propto v^2 / \sigma^2 \propto v$, i.e., the free energy for hydrating small hydrophobic solutes scales as solute volume, $v$.
Equation~\ref{eq:G_small} also provides a way to understand the temperature dependence of $\gcav$.
Garde et al.~\cite{Garde:PRL:1996} showed that $\sigma^2$ depends weakly on $T$, 
so that $\gcav \sim T \rho^2$, which increases with $T$ at ambient conditions.
These arguments provide a way to understand why the solubility of noble gases in water decreases with increasing temperature.
They also imply a negative entropy, $\scav$, for hydrating small hydrophobic solutes, 
which is consistent with the constrained hydrogen bonding network of their hydration waters.
%

For large volumes in bulk water, $\pvn$ remains Gaussian near its mean, 
but the likelihood of observing low-$N$ fluctuations is enhanced significantly (\fig{fig:hyd}{d})~\cite{Patel:JPCB:2010,Patel:JSP:2011}.
The enhanced low-$N$ fluctuations are a consequence of the proximity of water at ambient conditions 
to liquid-vapor coexistence, so that once a certain number of waters leave $v$, 
it becomes easier to further dewet the volume by nucleating a vapor bubble~\cite{Stillinger,LCW}.
The dewetting of large volumes is thus governed by interfacial physics, 
and the free energetic cost of lowering $N$ below a certain threshold is roughly $\gamma A$, 
where $\gamma$ is the water-vapor surface tension, and $A$ is the surface area of the vapor bubble, 
i.e., $- \ln \pvn \propto \gamma (v - N/\rho)^{2/3}$, where $(v - N/\rho)$ is the bubble volume.
Consequently, for large solutes, $\gcav \propto \gamma v^{2/3}$, and because $\gamma$ decreases with increasing $T$, so does $\gcav$.
In contrast with small solutes, $\scav$ is thus positive for large hydrophobic solutes~\cite{Huang:PNAS:2000}.
%

\subsection{Solute Size Dependent Crossover in Hydrophobic Hydration Free Energies}
Given the fundamental differences in the hydration of small and large hydrophobic solutes, 
at what solute size does the crossover occur from molecular solutes, 
whose hydration free energies scale with solute volume, 
to macroscopic solutes, whose hydration free energies scale with surface area?
Molecular simulations and theory have shown that the crossover occurs at a solute size of roughly 1~nm 
(\fig{fig:hyd}{e})~\cite{LCW,Huang:PNAS:2000,Huang:JPCB:2001,Rajamani:PNAS:2005}.
Interestingly, the building blocks of typical molecular assemblies, e.g., amino acid side chains, tend to be smaller than 1~nm, whereas the corresponding assemblies, e.g., folded proteins, are usually larger than 1~nm. 
Thus, to capture the driving forces for diverse self-assembly phenomena, 
e.g., micelle formation or protein folding, 
and to rationalize why those driving forces increase with temperature (sec.~5.1),
it is critical to understand the physics of hydrophobic hydration as it crosses over from one regime to another, 
and also be able to quantitatively characterize the corresponding free energies.
To this end, the physics of both Gaussian fluctuations and interface formation have been 
encoded in the Lum-Chandler-Weeks theory of hydrophobicity~\cite{LCW}, 
and in the theoretical contributions it has motivated~\cite{tenWolde:PRE:2002,Varilly:JCP:2011,Vaikuntanathan:PNAS:2016,Xi:PNAS:2016}.

\subsection{Experimental Evidence of Crossover in Hydrophobic Hydration}
Theory and molecular simulations played an important role in elucidating the differences 
between the hydration of small and large hydrophobic solutes;
however, experimental verification of these ideas was complicated by the fact that 
small non-polar solutes are only sparingly soluble in water, 
and their solubility decreases precipitously as solute size increases.
To address this challenge, Davis {\it et al.}~\cite{Davis:Nature:2012} 
studied the hydration shells of a family of linear alcohols, 
increasing in size from methanol to heptanol, 
over a wide range of temperatures from $0 - 100^\circ$C.
By using Raman scattering measurements with multivariate curve resolution, 
the authors found that the hydration shells of small alcohols displayed signatures of enhanced tetrahedral order, whereas the hydration shells of longer alcohols were relatively disordered, 
supporting the notion that the disruption of water structure 
near small and large hydrophobic solutes is fundamentally different.
In another creative study, Li and Walker~\cite{Li:PNAS:2011} used single-molecule force spectroscopy to investigate the hydration of several hydrophobic polymers with nanoscopic side-chains of different sizes. 
Although the hydrophobic polymers collapse and crash out of the aqueous solution, as expected,
the authors forcibly hydrated their monomers by using an AFM tip to pull on individual polymers (\fig{fig:hyd}{f}). 
By measuring the force required to unravel the polymers, 
the authors were then able to estimate the free energy of hydrating the monomers, $\ghyd$. 
For the smallest monomer studied, $\ghyd$ increased with temperature, 
whereas for the largest monomer, $\ghyd$ decreased above $50^\circ$C (\fig{fig:hyd}{g}).
These findings provide strong experimental support for the solute size dependent crossover in the physics of hydrophobic hydration, and confirm that the crossover length scale near ambient conditions is roughly 1~nm~\cite{Garde:PNAS:2011}.
%

\section{CHARACTERIZING HYDROPHOBICITY} 
A characterization of solute hydrophobicity, which quantifies how strongly water structure is perturbed by the solute, can inform the strength of hydrophobic interactions~\cite{Israelachvili:Nature:1996};
the more hydrophobic a solute, the more strongly it ought to bind other hydrophobic solutes.
Below, we first introduce canonical metrics of hydrophobicity before introducing modern measures, 
which are rooted in interfacial water density fluctuations, 
and can be used to estimate the hydrophobicity of heterogeneous, amphiphilic solutes (sec.~4), 
and inform their interactions and assemblies (sec.~5).
%

\subsection{Molecular vs Macroscopic Solutes}
Because the manner in which small and large solutes perturb water is fundamentally different, 
it stands to reason that their hydrophobicities should also be characterized differently.
The hydrophobicity of molecular solutes is typically characterized using solubility from the gas phase 
or partitioning from an oil phase, or equivalently, using the corresponding free energies, $\ghyd$ or $\gtr$, respectively; the larger the value of $\ghyd$, the more hydrophobic the solute (\fig{fig:char}{a})~\cite{Ben-Naim:JPC:1978,Ben-Naim:book}. 
As discussed in sec.~2.2, $\ghyd$ is closely related to the corresponding cavity hydration free energy, $\gcav$, 
which is informed by density fluctuations in bulk water.
Indeed, by recognizing that $\pvn$ in molecular volumes is Gaussian, and using it to estimate the $\gcav$ of n-butane conformers, Hummer et al.~\cite{Hummer:PNAS:1996} were able to show that the more compact cis conformation is favored over its trans conformation, in agreement with the corresponding estimates of $\ghyd$~\cite{Beglov:JCP:1994}.
Moreover, trends in $\ghyd$ can also shed light on the interactions and assemblies of molecular solutes, as we will discuss in sec.~5.1.
In contrast with small solutes, the hydrophobicity of large solutes (or surfaces) is determined by interfacial physics.
Below, we describe several different but related ways of characterizing hydrophobicity using interfacial physics;
in secs.~5.3~and~5.4, we will then illustrate how such characterizations inform the interactions between diverse extended solutes, ranging from flat, homogeneous surfaces to the rugged, heterogeneous surfaces of proteins.

\subsection{Water Droplet Contact Angles and Wetting Coefficients}
The hydrophobicity of a solid surface can be quantified by bringing a water droplet 
into contact with the surface and estimating the corresponding contact angle, $\theta$~\cite{Jiang:COChE:2019}.
A water droplet beads up to minimize contact with hydrophobic surfaces, 
whereas it spreads on hydrophilic surfaces (\fig{fig:char}{b}).
The contact angle is also related to the wetting coefficient, $k$, 
which quantifies the relative preference of a surface for liquid water over water vapor.
In particular, $k \equiv  (\gamma_{\rm SV} - \gamma_{\rm SL})/\gamma$, 
where $\gamma_{\rm SV}$ and $\gamma_{\rm SL}$ represent solid-vapor and solid-liquid surface tensions, respectively; interfacial physics stipulates that $k$ must be equal to $\cos\theta$~\cite{Jiang:SoftMatter:2019}.
Thus, the hydrophobicity of flat, homogeneous surfaces can be quantified using either $\theta$ or $k$.
However, such characterization cannot be performed for rugged, heterogeneous surfaces, such as those of proteins, which nevertheless use hydrophobic regions to interact with their binding partners~\cite{Granick:Science:2008}.
Once again, an understanding of water density fluctuations, this time in the vicinity of surfaces, 
provides the insights needed to address this challenge.

\subsection{Water Density Fluctuations at Extended Surfaces}
\begin{figure}[t]
\includegraphics[width=1.\textwidth]{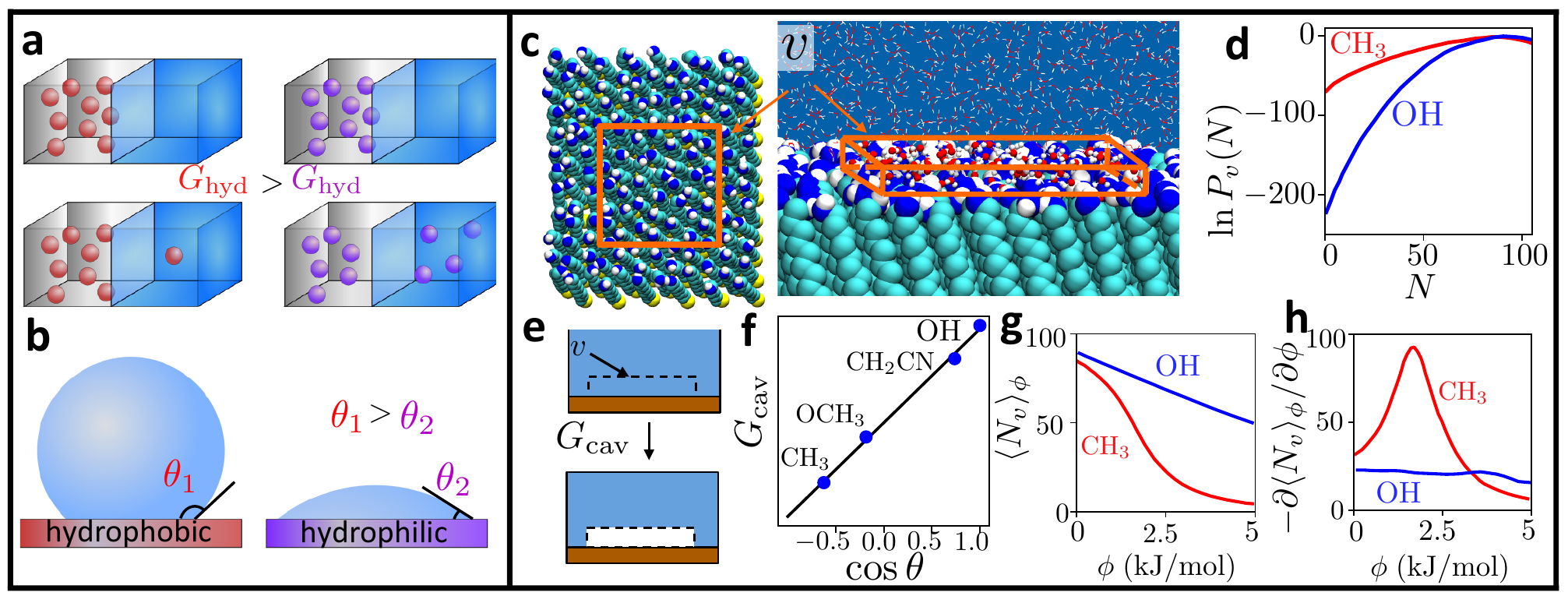}
\vspace{-0.05in}
\caption{
Characterizing hydrophobicity, water density fluctuations near surfaces, and interfacial dewetting. 
(a) The hydrophobicity of molecular hydrophobic solutes (red/purple spheres) 
is often characterized using their solubility in water or equivalently, their hydration free energy, $G_{\rm hyd}$.
(b) In contrast, the hydrophobicity of macroscopic surfaces is typically characterized using the water droplet contact angle, $\theta$;
water beads up on hydrophobic surfaces, but spreads on hydrophilic ones.
(c) Surface hydrophobicity is also reflected in the statistics of water density fluctuations, $\pvn$, 
in thin interfacial observation volumes; 
a cuboidal $v$ (orange) of width, 0.3~nm, and a cross-sectional area, 
$3\times3$~nm$^2$, is shown here adjacent to an OH-terminated self-assembled monolayer (SAM) surface. 
(d) Near the pol ar (OH-terminated) SAM surface, $\pvn$ is roughly Gaussian for all $N$,
whereas near the non-polar (CH$_3$-terminated) SAM surface, enhanced low-$N$ fluctuations are observed, reflecting the relative ease with which the hydrophobic surface can be dewetted~\cite{Patel:JPCB:2012}. 
(e, f) The free energy required to create a cavity, $\gcav$, next to a SAM surface
varies linearly with $\cos\theta$, and can be used to quantify surface hydrophobicity~\cite{Patel:PNAS:2011}. 
(g) In response to an unfavorable potential, $\phi N_v$, the average number of interfacial waters, $\langle N_v \rangle_\phi$ decreases linearly near the polar surface, 
whereas $\langle N_v \rangle_\phi$ decreases in a sigmoidal manner near the non-polar surface~\cite{Patel:JPCB:2012}.
(h) Correspondingly, the susceptibility, $-\partial \langle N_v \rangle_\phi / \partial \phi$ displays a marked peak near the non-polar SAM surface, 
suggesting that water near hydrophobic surfaces is situated at the edge of a collective dewetting transition, 
and is particularly susceptible to unfavorable perturbations~\cite{Patel:JPCB:2012}.
Panels d, g, and h adapted from Reference~\citenum{Patel:JPCB:2012} with permission; copyright 2012 American Chemical Society. 
Panel f adapted from reference~\citenum{Patel:PNAS:2011} with permission; copyright 2011 National Academy of Sciences.
}
\label{fig:char}
\end{figure}

To illustrate the relationship between surface hydrophobicity and interfacial water density fluctuations, 
we compare and contrast the statistics of water number fluctuations, $\pvn$, 
in thin interfacial volumes, $v$, adjacent to self-assembled monolayer (SAM) surfaces, 
terminated with either methyl (-CH$_3$) or hydroxyl (-OH) groups (\fig{fig:char}{c})~\cite{Patel:JPCB:2012}.
Although the average number of interfacial water molecules is similar in both cases, 
there are marked differences in the two $\pvn$-curves at low-$N$ (\fig{fig:char}{d}).
For the non-polar CH$_3$-terminated SAM surface, the likelihood of low-$N$ fluctuations is enhanced significantly relative to the polar OH-terminated SAM surface.
Thus, the free energetic cost of displacing a sufficient fraction of interfacial waters is lower near a hydrophobic surface than a hydrophilic surface.
In other words, the ease with which surface-water interactions can be disrupted, by dewetting interfacial volumes, depends on how weak those interactions were to begin with.
%

%
%

\subsection{Interfacial Cavity Hydration Free Energy as a Measure of Hydrophobicity}
Because dewetting hydrophobic surfaces is easier than hydrophilic ones, 
the free energy required to create a cavity near a surface, $\gcav$, 
can serve as a measure of surface hydrophobicity.
Moreover, $\gcav$ can also be related to conventional measures of hydrophobicity, such as $\theta$, 
using interfacial physics.
In particular, $\gcav \approx  (\gamma + \gamma_{\rm SV} - \gamma_{\rm SL}) A_{\rm c}$, where $A_{\rm c}$ is the cross-sectional area of $v$ (\fig{fig:char}{e});
Young's equation then suggests a linear relationship between $\gcav$ and $\cos\theta$, i.e., $\gcav \approx (1 + \cos\theta)A_{\rm c}$.
Patel et al.~\cite{Patel:PNAS:2011} estimated $\gcav$ for a family of SAM surfaces with different end-groups, which confer the surfaces with a wide range of hydrophobicities, and found that $\gcav$ indeed varies linearly with $\cos\theta$ (\fig{fig:char}{f}).
Thus, the hydrophobicity of flat surfaces can be equivalently characterized using either $\theta$ or $G_{\rm cav}$. 
Importantly, the connection between $\gcav$ and $\cos\theta$ suggests that 
even when $\theta$ is ill-defined, e.g., for complex surfaces, such as those of proteins,
$\gcav$ can nevertheless be used to characterize hydrophobicity~\cite{Godawat:PNAS:2009,Patel:PNAS:2011}. 
In addition to enhanced interfacial water density fluctuations, and the corresponding ease of cavity creation, 
the disruption of water structure near hydrophobic surfaces is also reflected in other interfacial properties, such as isothermal compressibility, transverse density correlations, distribution of water dipole orientations, interfacial diffusivity, orientational relaxation times, hydrodynamic slip and Kapitza resistance~\cite{Mittal:PNAS:2008,Sarupria:PRL:2009,Godawat:PNAS:2009,Lee:JCP:1984,Laage:ACR:2012,Shenogina:PRL:2009,Adjari:PRL:2006}.
Like $\gcav$, some of these quantities have also been employed as reporters of surface hydrophobicity, and used to interrogate surfaces with nanoscale complexity~\cite{Giovambattista:JPCC:2007,Giovambattista:PNAS:2008,Acharya:Faraday:2010,Heyden:PRL:2013,Fogarty:2014,Shin:JPCB:2018,Monroe:PNAS:2018,Heyden:WIREs:2019}.
%

\subsection{Response of Interfacial Waters to Unfavorable Perturbations}
In addition to serving as descriptors of surface hydrophobicity, 
the rare low-$N$ tails of interfacial water density fluctuations (\fig{fig:char}{d}) 
also dictate the response of interfacial waters to unfavorable perturbations, 
e.g., interfacial dewetting facilitated by the approach of a binding partner (secs.~6.3, 6.4)~\cite{Setny:PNAS:2013,Mondal:PNAS:2013}, the addition of a co-solvent~\cite{Xiao:Langmuir:2017}, etc.
In particular, consider the contrasting response of water near hydrophobic and hydrophilic surfaces 
to an unfavorable linear potential, $\phi N_v$, where $\phi$ represents the potential strength, and $N_v$ is the number of waters in the interfacial volume, $v$.
As $\phi$ is increased, the average number of waters, $\langle N_v \rangle_\phi$, near the OH-terminated SAM surface decreases gradually (linear response); in contrast, near the CH$_3$-terminated SAM surface, $\langle N_v \rangle_\phi$ displays a sharp decrease (\fig{fig:char}{g})~\cite{Patel:JPCB:2012,Xi:JCTC:2016}.
Correspondingly, the susceptibility, $-\partial \langle N_v \rangle_\phi / \partial \phi$, displays a peak near the hydrophobic surface (\fig{fig:char}{h}).
Such a peak in susceptibility is a well-known feature of phase transitions, 
and is a direct consequence of enhanced low-$N$ fluctuations~\cite{Patel:JPCB:2012}.
Thus, low-$N$ fat tails in the statistics, $\pvn$, of interfacial water density fluctuations 
situate interfacial waters at the edge of a dewetting transition, 
and render them sensitive to unfavorable perturbations.
Interestingly, the susceptibility of interfacial waters to unfavorable perturbations 
also manifests itself in the remarkable sensitivity of $\gcav$ to the characteristics of a hydrophobic surface.
Indeed, as we will see in sec.~4, small differences in surfaces cues, 
such as curvature, substitution of individual chemical groups (e.g., mutations), 
or rearrangement of moieties (patterning), can all substantially modulate $\gcav$, 
and thereby surface hydrophobicity.
In contrast, hydrophilic surfaces tend to be relatively insensitive to subtle changes in surface properties~\cite{Acharya:Faraday:2010}.

\section{HETEROGENEOUS SOLUTES: HYDROPHOBICITY IN CONTEXT} 
The most interesting solutes, which participate in hydrophobic interactions, tend to be amphiphilic; 
polar regions facilitate their hydration, whereas non-polar regions facilitate their interactions. 
The extent to which water structure is perturbed by such heterogeneous solutes 
depends on their detailed chemical and topographical patterns
in ways that can be both surprising and non-intuitive.

\subsection{Small Amphiphilic Solutes} 
\begin{figure}[h]
	\includegraphics[width=0.9\textwidth]{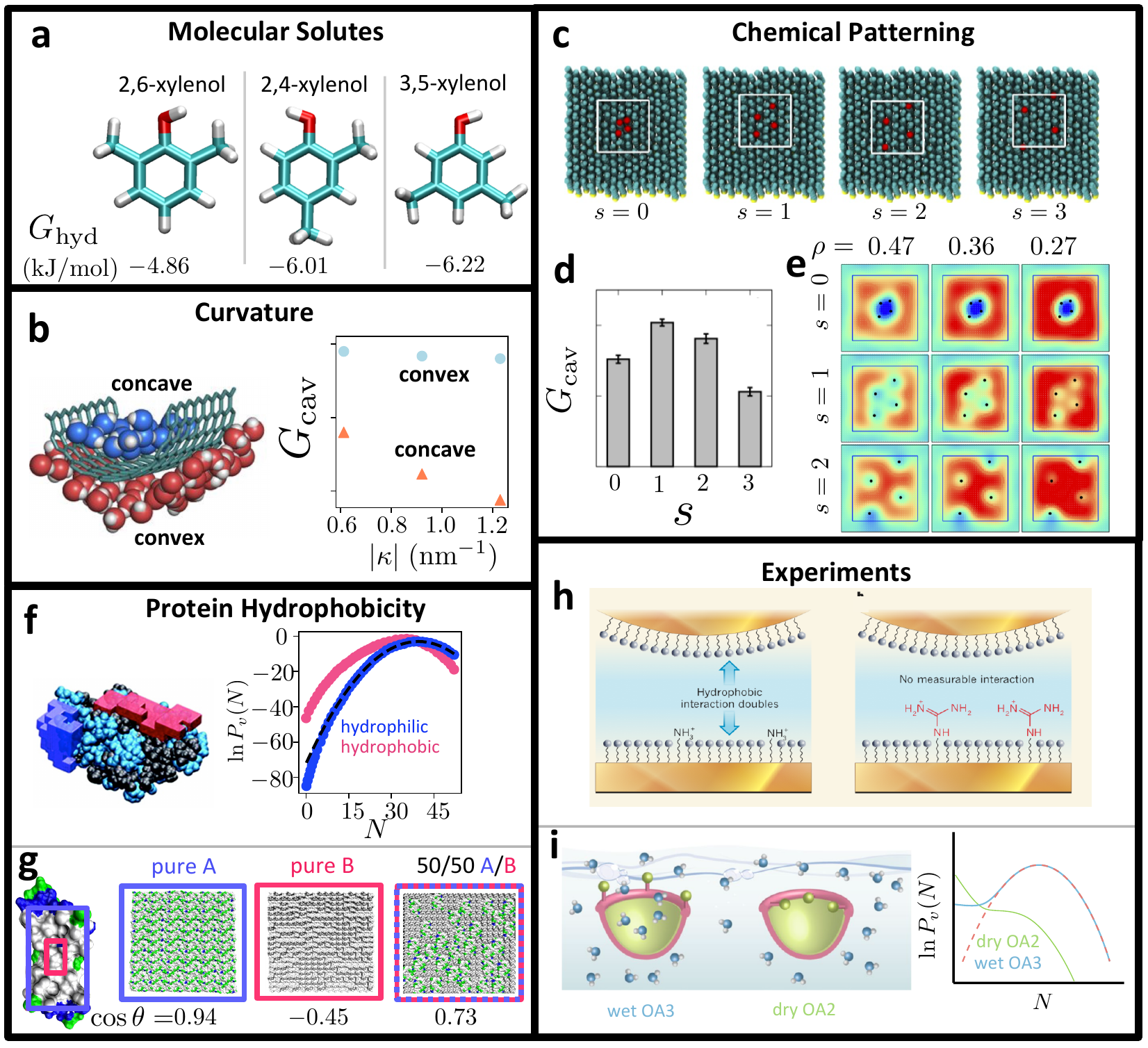}
	\caption{
		Context-dependent hydrophobicity of heterogeneous solutes and surfaces. 
		(a) The hydrophobicity of xylenol isomers, as quantified by their hydration free energies, 
		depends on the arrangement of their chemical groups with respect to one another~\cite{Fennell:JSP:2011}. 
		(b) The hydrophobicity of hemi-cylindrical surfaces, as quantified by the free energetic cost of creating interfacial cavities, $\gcav$, depends on their curvature; concave surfaces are more hydrophobic than convex ones~\cite{Xi:PNAS:2017}.
		(c, d) For a family of SAM patches containing four -OH groups (magenta) separated by $s \in [0,3]$~CH$_3$ groups (cyan), $\gcav$ varies non-monotonically with $s$. 
		(e) As water is displaced from the observation volume, $v$ (white), wet (blue) and dry (red) patches appear in $v$, and are shown for select partially dewetted states ($\rho$ is the normalized water density in $v$)~\cite{Xi:PNAS:2017}. 
		(f) The statistics of interfacial water density fluctuations, $\pvn$, 
		near a hydrophobic patch (red) on the BphC enzyme (cyan/black) display fat low-$N$ tails, 
		whereas $\pvn$ near a hydrophilic protein patch (blue) is approximately Gaussian~\cite{Patel:JPCB:2012}.
		(g) Wang et al.~\cite{Wang:PNAS:2011} constructed mixed surfaces using hydrophilic (blue, A) and hydrophobic (red, B) regions of the melittin protein, and found that the mixed surfaces were substantially more hydrophilic than expected by additivity.
		(h) Ma et al.~\cite{Abbott:Nature:2015} demonstrated that the attraction between a hydrophobic AFM tip and 
		an amphiphilic SAM surface increased in strength when polar amine groups were replaced with 
		charged ammonium groups, but disappeared altogether when they were replaced by 
		similarly charged guanidinium groups.
		(i) Barnett et. al~\cite{Gibb:Nature:2020} showed that the orientation of methyl groups (green spheres) 
		on otherwise identical octa-acid cavitands, OA3 and OA2, had a dramatic effect on their hydration, 
		stabilizing either a wet or dry state.
		Panel a adapted from Reference~\citenum{Fennell:JSP:2011} with permission. 
		Panels b, c, d and e adapted from Reference~\citenum{Xi:PNAS:2017} with permission; copyright 2017 National Academy of Sciences. 
		Panel f adapted from Reference~\citenum{Patel:JPCB:2012} with permission; copyright 2012 American Chemical Society. 
		Panel g adapted from Reference~\citenum{Wang:PNAS:2011} with permission; copyright 2011 National Academy of Sciences. 
		Panel h adapted from Reference~\citenum{Garde:Nature:Abbott} with permission. 
		Panel i adapted from Reference~\citenum{Garde:Nature:Gibb} with permission.
	}
	\label{fig:context}
\end{figure}

The hydrophobicity of a small amphiphilic solute, as quantified by $\ghyd$, depends not just on 
the chemical groups present in the solute, but also on their spatial positions with respect to one another.
For example, consider three different xylenol molecules, highlighted by Fennel and Dill~\cite{Fennell:JSP:2011},
which have the exact same number and types of chemical groups; 
the solutes differ only in the locations of their methyl groups (\fig{fig:context}{a}).
Both common intuition and more involved group additivity approaches would predict that these three compounds have similar, if not identical, hydration free energies. 
Instead, experiments show that 3,5-xylenol is more than twice as soluble in water as 2,6-xylenol.
This example highlights that the extent to which a solute perturbs water structure, 
and therefore its hydrophobicity, can depend sensitively on the chemical and topographical 
cues that the solute presents to its hydration waters. 
Such sensitivity has motivated approaches 
which make use of the inhomogeneous water density distribution near a solute to approximately estimate its $\ghyd$~\cite{WaterMap,Gilson:JCP:2012,Huggins:JPCB:2013,Levy:JPCB:2017}. 

\FloatBarrier
%
%
\FloatBarrier

\subsection{Curved and Chemically Patterned Surfaces}
As discussed in sec.~3.4, the hydrophobicity of heterogeneous surfaces, which display nanoscale curvature and/or chemical patterning, can be characterized using the free energy, $\gcav$, of creating an interfacial cavity. 
Xi {\it et al.}~\cite{Xi:PNAS:2017} quantified $\gcav$ near both concave and convex hemi-cylindrical surfaces 
with nanoscale curvatures, and found that interfacial waters could be displaced to form cavities more readily near concave surfaces than near convex surfaces of the same curvature (\fig{fig:context}{b}).
Thus, concave non-polar surfaces are more hydrophobic than convex ones.
The authors also found that concave and convex curvatures do not influence hydrophobicity in similar ways;
the hydrophobicity of concave surfaces increased with curvature, 
whereas that of convex surfaces was relatively insensitive to curvature.
Along these lines, Mittal and Hummer~\cite{Mittal:Faraday:2010} found that rough surfaces, 
which possess both concave and convex curvatures, tend to be more hydrophobic than flat ones.
%

%
Xi et al.~\cite{Xi:PNAS:2017} also studied chemically patterned SAM patches containing a small fraction of polar groups in an otherwise non-polar background; varying the separation, $s$, 
between adjacent polar groups led to four different patterns (\fig{fig:context}{c}).
The authors found that surface hydrophilicity, as quantified by $\gcav$, 
was not only different for the different surfaces, it varied non-monotonically with $s$ (\fig{fig:context}{d}).
A visualization of the corresponding dewetting pathways provided a striking illustration of the 
collective response of the SAM hydration waters to the chemical context presented by the different patches (\fig{fig:context}{e}).
Along these lines, Acharya et al.~\cite{Acharya:Faraday:2010} found that adding a polar group to a non-polar surface suppresses its hydrophobicity substantially, but adding a non-polar group to a polar surface does not correspondingly enhance surface hydrophobicity.
Similarly, Giovambattista et al.~\cite{Giovambattista:PNAS:2009} found that inverting the polarity of silica surfaces modulates their hydrophobicity. 
These examples highlight that seemingly similar surfaces with identical chemical compositions, 
but with subtle variations in chemical patterning and/or topography, can display substantial differences in their aversion for water.

\subsection{Protein Hydrophobicity}
Protein surfaces display nanoscale chemical and topographical patterns, 
which perturb water structure in complex ways that are not easy to anticipate~\cite{Zhou:Science:2004,Liu:Nature:2005}.
Patel et al.~\cite{Patel:JPCB:2012} showed that, in spite of these complexities, 
water density fluctuations can be used to characterize the hydrophobicity of heterogeneous protein surfaces.
In particular, the authors showed that $\pvn$ can display fat low-$N$ tails near hydrophobic protein regions, 
and tends to be Gaussian near hydrophilic regions (\fig{fig:context}{f}).
Similarly, Rego et al.~\cite{Rego:JACS} found that the hydration shells of diverse proteins were susceptible to unfavorable perturbations, suggesting that protein hydrophobicity ought to be sensitive to its chemical and topographical context.
Indeed, Xi et al.~\cite{Xi:PNAS:2017} showed that the hydrophobicity of a protein patch, as quantified by $\gcav$, 
can be modulated substantially by swapping the positions of two residues in the patch.
Similarly, Wang et al.~\cite{Wang:PNAS:2011} studied mixed surfaces comprised of polar and non-polar regions of the melittin protein, and found that polar patches disproportionately suppress surface hydrophobicity (\fig{fig:context}{g}).
Finally, Rego et al.~\cite{Rego:PNAS:2021} classified protein regions according to their ease of dewetting, 
and found that hydrophobic and hydrophilic patches had remarkably similar chemical compositions, 
and contained comparable numbers of non-polar and polar atoms.

\subsection{Experimental Demonstrations of Context-Dependent Hydrophobicity}
Ma et al.~\cite{Abbott:Nature:2015} characterized the hydrophobicity of heterogeneous surfaces by measuring the adhesive force between a hydrophobic atomic force microscopy (AFM) tip and the surface of interest;
the stronger the force, the more hydrophobic the surface.
By studying binary SAM surfaces, composed of a 60/40 mixture of -CH$_3$ and cationic end-groups, the authors showed that surface hydrophobicity displayed a mystifying dependence on both the nature of the cationic groups and their charge (\fig{fig:context}{h}).
Surprisingly, surfaces with cationic ammonium ions were found to be nearly twice as hydrophobic as surfaces with uncharged amine groups.
Even more surprisingly, surfaces with the cationic guanidinium ions were found to be hydrophilic, i.e., they were not attracted to the hydrophobic AFM tip.
In another interesting study, Barnett et al.~\cite{Gibb:Nature:2020} studied synthetic nanoscale bowl-shaped hosts or cavitands, which are lined with hydrophilic acid groups on the outside to make them water soluble, and hydrophobic groups on the inside.
By studying variants of the octa-acid (OA) cavitands, which differ only in the orientation of the methyl groups at the rim of the bowl, i.e., the methyls point inwards (OA2) or upwards (OA3), 
the authors found that OA3 bowl was wet, whereas the OA2 bowl was dry (\fig{fig:context}{i}).
Molecular simulations shed further light on these interesting observations, 
highlighting that the presence of a fat low-$N$ tail in the $\pvn$ for OA3, 
which suggests that although OA3 is wet, it is situated at the edge of a dewetting transition.
Presumably, the subtle change in the orientation of the methyl groups 
in going from OA3 to OA2 was then sufficient to trigger dewetting. 

\subsection{Shortcomings of Surface Area Models and Additive Approaches}
The importance of hydrophobic interactions in driving diverse self-assembly processes has inspired 
numerous schemes for the efficient, but approximate characterization of solute hydrophobicity.
For example, surface area (SA) models~\cite{Eisenberg:Nature:1986,Roux:BC:1999,Kang:JPC:1987}
assume that the hydrophobicity of an exposed non-polar patch is proportional to its surface area.
However, the hydration of small hydrophobic solutes is not governed by interfacial physics, 
but by the entropic constraints on their hydration waters (sec.~2.2). 
Even for extended solutes, hydrophobicity can depend on surface curvature, 
particularly for concave surfaces (sec.~4.2). 
These shortcomings of SA models necessitate the use of an effective surface tension, 
whose value can depend sensitively on the class of solutes used to estimate it~\cite{Mobley:JCTC:2009}.
Another popular class of approximate approaches for characterizing protein hydrophobicity uses a divide-and-conquer strategy.
Such approaches employ hydropathy scales (or scoring functions)~\cite{Kyte:JMB:1982,Ferrara:JMC:2004,Bonella:JPCB:2014,Eisenberg:ARB:1984,Rose:ARBBS:1993}, which assign an index (or score) to an amino acid residue based on some measure of its aversion to water (e.g., its transfer free energy from oil); 
the hydrophobicity of a protein patch is then estimated as a sum of the hydrophobicities of constituent amino acids.
Numerous hydropathy scales exist, and they can be quite different from one another~\cite{Cornette:JMB:1987}. 
Importantly, they tend to fare poorly in predicting the driving forces for hydrophobic interactions and assemblies~\cite{Kortemme:PNAS:2002,Kollman:ACR:2000,Kister:PNAS:2008}. 
The failure of such additive approaches does not stem from an imperfect hydropathy scale or scoring function, 
but rather from the underlying assumption that each residue has a unique hydrophobicity. 
As the examples provided in this section highlight, the collective and non-trivial response of 
protein hydration waters to the chemical and topographical cues presented by a protein surface 
means that the hydrophobicity of a residue is not unique, but depends sensitively on the context within which it appears on the protein surface. 

\section{HYDROPHOBIC INTERACTIONS AND ASSEMBLIES: DRIVING FORCES}
The association of hydrophobic solutes, which minimizes their disruption of water structure, 
forms the basis for diverse processes, ranging from micelle formation and protein folding, 
to biomolecular interactions and colloidal assemblies~\cite{Shen:BJ:2006,Palmer:JPCL:2012,Thirumalai:ACR:2012,Vashisth:Proteins:2013,Remsing:JPCB:2018,Whitesides:Science:2002}.
The thermodynamic driving forces for such interactions and assemblies depend on the extent to which 
the solutes disrupt water structure (relative to their assemblies), which in turn, depends sensitively on solute (and assembly) characteristics, such as size, shape, curvature and chemical patterning (sec.~4). 
Indeed, as described below, differences in the hydration of small and large hydrophobic solutes (sec.~2) alone 
can spawn a veritable menagerie of assembly physics. 
%

\vspace{-0.15in}
\subsection{Assemblies of Small Solutes}
Molecular self-assembly often involves sub-nm building blocks, 
which come together to form assemblies that are larger than 1~nm in size, 
e.g., the association of amphiphilic surfactant molecules to form micelles, 
or the folding of proteins into their native structures (\fig{fig:thermo}{a}).
Due to their different sizes, molecular solutes and their assemblies perturb water structure differently.
Thus, the hydration free energies, $G_{\rm hyd}$, of individual solutes scale with their volumes, 
whereas the hydration free energy of the assembly scales as its surface area (sec.~2).
The driving force for assembling of $n$ small solutes into a large aggregate is then:
$\Delta G_{\rm agg} \approx c \gamma n^{2/3} - n \ghyd$, 
where $c$ is a constant that depends on aggregate shape.
For sufficiently large $n$, the second term can dominate, so that 
$\Delta G_{\rm agg}$ depends primarily on solute hydrophobicity, as quantified by $\ghyd$.
For example, $\ghyd$ for surfactant molecules increases linearly 
with the length of their non-polar tails~\cite{Chandler:lecture};
correspondingly, the driving force for micelle formation increases with tail length, 
and results in an exponential decrease in the critical micelle concentration (CMC).
%

%
The differences in how small solutes and their assemblies perturb water structure are also responsible for 
the peculiar strengthening of hydrophobic interactions with increasing temperature.
In particular, as temperature is increased, $\ghyd$ increases, whereas $\gamma$ decreases (sec.~2);
consequently, the difference between the two, which represents the driving force for assembly, 
increases with temperature~\cite{Chandler:Nature:2005}.
Indeed, experiments on a variety of surfactants have shown that, at ambient conditions, 
their CMC goes down as temperature is increased~\cite{fig4:cmc}, confirming that micelle formation is favored at higher temperatures (\fig{fig:thermo}{b}).
Similarly, the stability of folded proteins also increases with temperature near ambient conditions~\cite{Tanford:APC:1968,Privalov:CRBMB:1990,Kim:PNAS:2016}.
As temperature is increased further, protein stability eventually decreases 
as the conformational entropy gain upon unfolding prevails over 
the entropy loss due to hydrophobic hydration. 
Thus, folded proteins undergo denaturation upon both cooling and heating, albeit for very different reasons (\fig{fig:thermo}{c})~\cite{Matysiak:JPCB:2012}.
%

\vspace{-0.1in}
\subsection{Binding of Small Solutes to Large Solutes}
\begin{figure}[h]
\includegraphics[width=1.\textwidth]{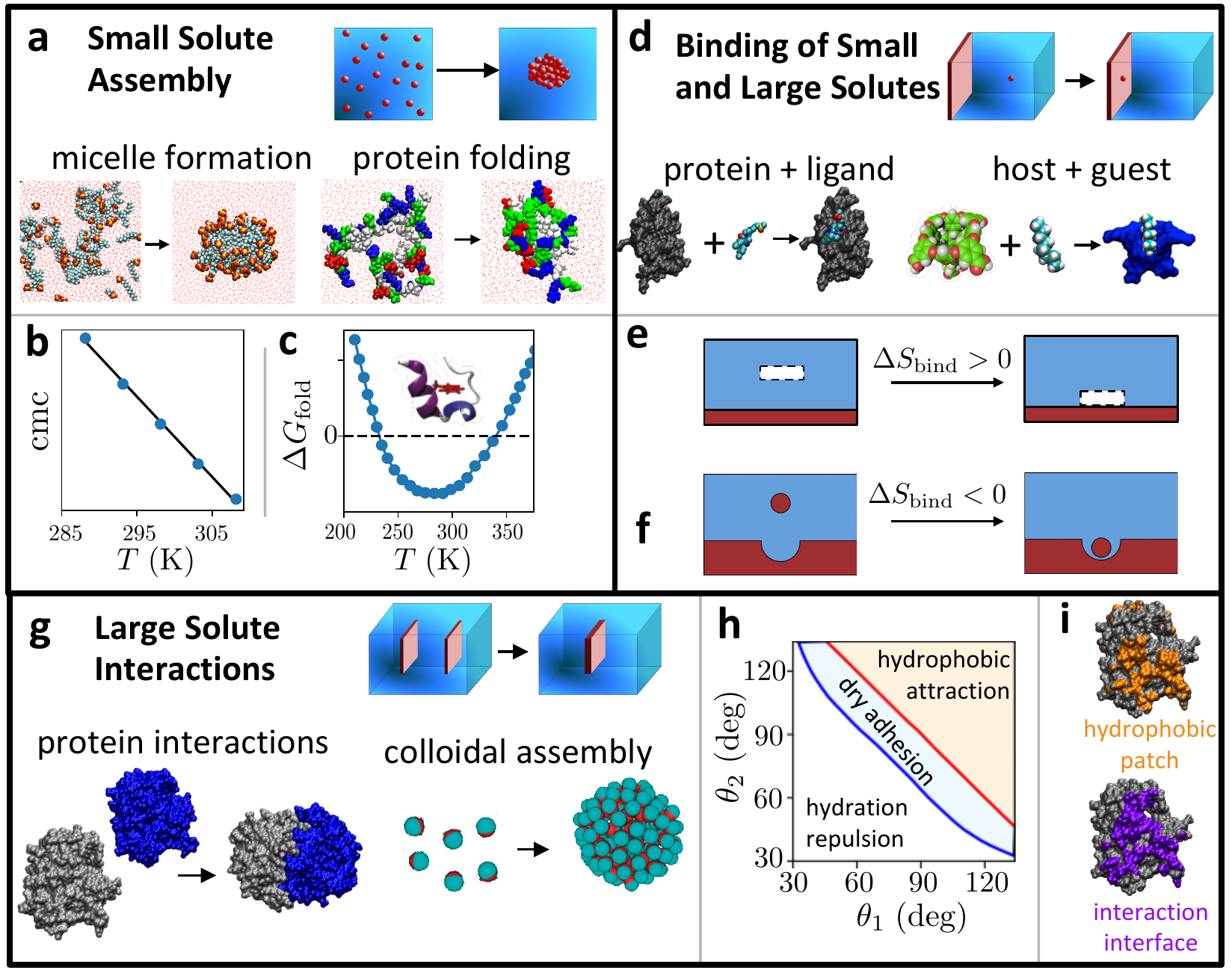}
\caption{
Thermodynamic forces that drive hydrophobic interactions and assemblies. 
(a) The association of small, hydrophobic solutes plays a role in micelle formation (left) and protein folding (right).  
Near ambient conditions, the hydrophobic driving force becomes stronger with increasing temperature, 
leading to a decrease in 
(b) the critical micelle concentration of dodecyl trimethyl ammonium chloride~\cite{fig4:cmc} and 
(c) the folding free energy of the Trp-Cage mini-protein~\cite{Kim:PNAS:2016}.
(d) Protein-ligand binding (left) and host-guest interactions (right) involve the association of small hydrophobic solutes with extended surfaces~\cite{Gibb:Nature:2020,Tang:JPCB:2017}. 
(e) The binding of a hard solute (or cavity) to a hydrophobic surface is favored by entropy~\cite{Patel:PNAS:2011}, whereas
(f) the binding of methane to a concave pocket is opposed by entropy~\cite{Setny:JCTC:2010}.
(g) Interactions between extended hydrophobic solutes play an important role in protein interactions (left) and colloidal assemblies (right)~\cite{fig4:colloid}.
(h) Kandu{\v c} and Netz~\cite{Netz:PNAS:2015} studied the interactions between extended surfaces with widely varying polarities, and found that hydrophobic surfaces were attracted to one another, whereas hydrophilic surfaces experienced hydration repulsion; a weakly attractive regime was observed for intermediate polarities.
(i) Rego et al.~\cite{Rego:PNAS:2021} showed that the most hydrophobic patch of the thymidylate synthase protein, which dewets readily in response to an unfavorable potential, has significant overlap with its interaction interface.
Panel b adapted from Reference~\citenum{fig4:cmc} with permission; copyright 2005 Elsevier. 
Panel c adapted from Reference~\citenum{Kim:PNAS:2016} with permission; copyright 2016 National Academy of Sciences. 
Panel d (right) adapted from References~\citenum{Tang:JPCB:2017,Gibb:Nature:2020} with permission; copyright 2017 American Chemical Society. 
Panel g (right) adapted from Reference~\citenum{fig4:colloid} with permission; copyright 2015 American Institute of Physics. 
Panel h adapted from Reference~\citenum{Netz:PNAS:2015} with permission; copyright 2015 National Academy of Sciences. 
Panel i adapted from Reference~\citenum{Rego:PNAS:2021} with permission; copyright 2021 National Academy of Sciences.
}
\label{fig:thermo}
\end{figure}

The interactions between sub-nm solutes and those that are larger form the basis for numerous interesting phenomena, including protein-ligand binding and supramolecular chemistry (\fig{fig:thermo}{d}).
When a small non-polar solute binds a hydrophobic surface, water can no longer hydrogen bond around the solute; binding is thus accompanied by the release of constrained hydration waters, and ought to be favored by entropy (\fig{fig:thermo}{e})~\cite{Patel:PNAS:2011}.
However, the thermodynamic signatures of such processes can be sensitive to the binding context~\cite{Baron:ARPC:2013,Monroe:PNAS:2021}.
For example, Setny et al.~\cite{Setny:JCTC:2010} found that the binding of methane to 
a concave pocket was favored by enthalpy, rather than entropy.
In particular, the release of methane's hydration waters didn't drive binding as much as the dewetting of the pocket (\fig{fig:thermo}{f}).
Indeed, the favorable dewetting of the pocket is consistent with the observation that concave non-polar regions are particularly hydrophobic (sec.~4.2, \fig{fig:context}{b})~\cite{Xi:PNAS:2017}.
A series of experimental studies, which employed isothermal titration calorimetry to 
characterize the thermodynamics of supramolecular host-guest interactions~\cite{Suating:CS:2020} 
and protein-ligand binding~\cite{Snyder:PNAS:2011}, 
have similarly highlighted that the affinity of small and large solutes for one another, 
and in particular, whether binding is driven by enthalpy or entropy, 
can be remarkably sensitive to small changes in the properties of either solute.
%

\FloatBarrier
%
\FloatBarrier

\vspace{-0.1in}
\subsection{Interactions Between Extended Solutes}
The favorable interactions between extended hydrophobic surfaces drive diverse biomolecular and colloidal assemblies~\cite{Israelachvili:Nature:1996} (\fig{fig:thermo}{g}). 
In contrast with molecular solutes, the driving force for assembling extended surfaces is governed by interfacial physics, and decreases with increasing temperature~\cite{Patel:PNAS:2011}.
Moreover, when two surfaces bind, not only are surface-water interactions disrupted, they are replaced by direct interactions between the surfaces.
The strength of the former depends on surface hydrophobicity, as quantified by $\theta$ (sec.~3.2) or $\gcav$ (sec.~3.4), but the latter can also influence the overall interactions between surfaces.
To clarify the interplay between such solvent-mediated and direct interactions, 
Kandu{\v c} and Netz~\cite{Netz:PNAS:2015} quantified the overall binding strength between 
two planar surfaces spanning a wide range of polarities (\fig{fig:thermo}{h}).
The authors found that the binding of non-polar surfaces is highly favorable, and is driven by their unfavorable interactions with water.
Conversely, the binding of both moderate and strongly polar surfaces was found to be unfavorable in spite of the favorable direct interactions between the surfaces; it turns out that the surfaces interact more favorably with water than with one another, which gives rise to a net repulsion between them.
Finally, the authors observed favorable binding between nearly neutral-wetting surfaces over a narrow range of surface polarities; in such a dry adhesion regime, favorable direct interactions between the surfaces are able to edge out favorable surface-water interactions.
Interestingly, binding driven by unfavorable surface-water interactions (hydrophobic attraction) 
spans a much larger range of surface polarities than binding driven by favorable direct interactions between the surfaces (dry adhesion).

\vspace{-0.1in}
\subsection{Protein Hydrophobicity Informs Protein Interaction Interfaces}
Protein--protein interactions play a crucial role in numerous biological processes, 
and are believed to be driven, at least in part, by hydrophobic interactions~\cite{Thirumalai:ACR:2012}.
Indeed, the findings of Kandu{\v c} and Netz~\cite{Netz:PNAS:2015} suggest that identifying hydrophobic protein regions should provide a promising approach for uncovering the interfaces through which proteins interact~\cite{Young:PS:1994,Keskin:CR:2008}. 
However, identifying hydrophobic protein patches is challenging because protein surfaces are heterogeneous, 
and their hydrophobicity depends in non-trivial ways on the nanoscale chemical and topographical patterns they display~\cite{Rossky:Faraday:2010,Cui:Bioinformatics:2014} (sec.~4). 
To address this challenge, Rego et al.~\cite{Rego:PNAS:2021} applied an unfavorable potential to all the protein hydration waters; the protein regions that nucleated cavities most readily in response to the potential were then deemed as the most hydrophobic.
By comparing these protein regions against experimentally-determined protein-protein interaction interfaces, the authors further showed that the most hydrophobic protein patches were also likely to mediate their interactions (\fig{fig:thermo}{i}); 
such a correspondence is quite remarkable, and points to the importance of hydrophobicity in driving protein interactions.
%

\section{DEWETTING AND BARRIERS TO ASSEMBLY}
For non-polar solutes in water to self-assemble, solute-water interactions must be disrupted.
Barriers to assembly thus depend on the ease of dewetting the region between hydrophobic solutes.
Solute (de)hydration thus dictates not just the thermodynamics, 
but also the mechanistic pathways and kinetics of assembly~\cite{Vembanur:JPCB:2013,Tiwary:PNAS:2015}.

\subsection{Barriers to Small Solute Assemblies}
The potential of mean force for the association of small hydrophobic solutes, 
which represents the dependence of free energy on the separation between solutes, displays two minima: 
a stable contact minimum and a metastable solvent-separated minimum, 
which are separated by a desolvation barrier (\fig{fig:kinetics}{a})~\cite{Vembanur:JPCB:2013,Pangali:JCP:1979}.
Because the assembly of small hydrophobic solutes is driven by the release of their constrained hydration waters (sec.~5.1), the contact minimum is favored by entropy~\cite{Ghosh:JCP:2002,Koga:PCCP:2003}.
In contrast, the solvent-separated minimum is favored by enthalpy and disfavored by entropy, 
due to the highly constrained hydrogen bonding network of the hydration waters that are shared by the solutes.
Many of these constrained hydration waters are released as the system approaches the desolvation barrier, 
so the barrier configuration is favored by entropy, and its height decreases with increasing temperature.
Desolvation barriers can also hinder the favorable interactions between well-hydrated hydrophobic sub-units of flexible molecules (e.g., polymer side-chains or protein residues), which can drive the conformational transitions of such macromolecules (e.g., polymer collapse or protein folding).
Indeed, mechanistic studies of the coil-to-globule transition have shown that the bottleneck to 
hydrophobic polymer collapse is the formation and dewetting of a sufficiently large non-polar cluster 
with water density fluctuations playing an important role in stabilizing the cluster~\cite{tenWolde:PNAS:2002,Miller:PNAS:2007,Dhabal:JPCB:2021} (\fig{fig:kinetics}{b}).

\subsection{Assembly at Hydrophobic Surfaces is Barrierless}
\begin{figure}[h]
\vspace{-0.1in}
\includegraphics[width=1.\textwidth]{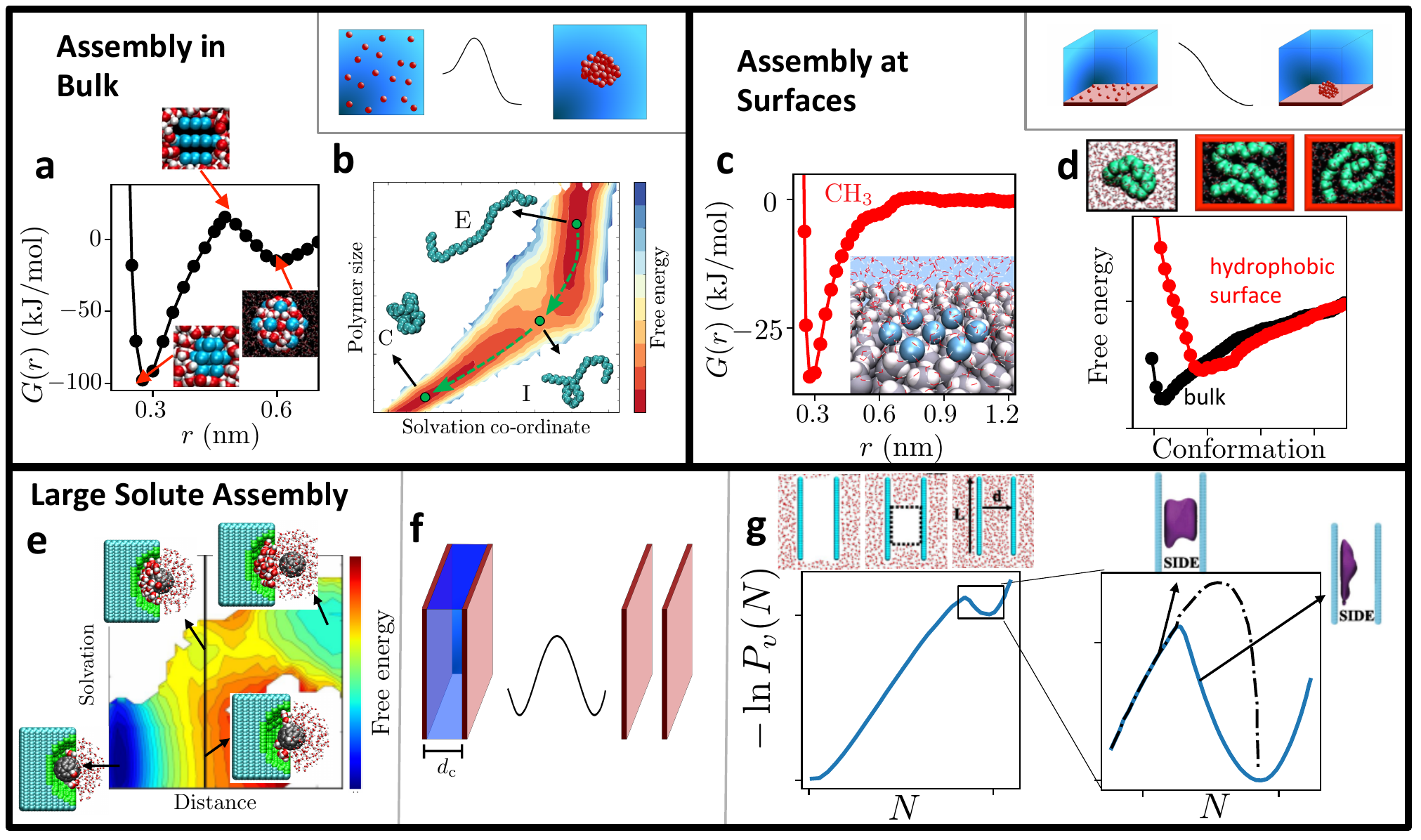}
\vspace{-0.1in}
\caption{
Desolvation barriers to assembly. 
(a) The potential of mean force for the association of small hydrophobic solutes (cyan) in water (red/white)
features a stable contact minimum, a metastable solvent separated minimum, and a desolvation barrier separating the two basins~\cite{Vembanur:JPCB:2013}.
(b) The free energy landscape corresponding to the collapse of a hydrophobic polymer in water is shown 
as a function of the number of waters in its hydration shell, $N_v$, and the polymer radius of gyration, $R_{\rm g}$;
the landscape highlights that an intermediate (I), partially dewetted configuration represents the barrier to the transition from an extended (E) to a collapsed (C) state~\cite{Dhabal:JPCB:2021}. 
(c) In contrast with the assembly in bulk water, the assembly of small hydrophobic solutes 
at an extended hydrophobic surface is barrierless~\cite{Vembanur:JPCB:2013}.
(d) Consequently, a hydrophobic polymer adsorbed to a hydrophobic surface displays enhanced conformational fluctuations~\cite{Jamadagni:JPCB:2009}.
(e) The free energy landscape of fullerene binding to a non-polar cavity is shown as a function of the separation between the binding partners and the solvation of the cavity; 
slow solvent fluctuations between wet and dry cavity states (seen at intermediate separations) 
impede the kinetics of assembly~\cite{Tiwary:PNAS:2015,Mondal:JSP:2020}.
(f) Although water confined between extended hydrophobic surfaces becomes metastable below a critical separation, $d_{\rm c}$, dewetting can be hindered by large free energy barriers at separations well below $d_{\rm c}$. 
(g) Dewetting in confinement between extended hydrophobic surfaces (cyan) is facilitated by water density fluctuations, which nucleate isolated interfacial cavities (purple) adjacent to one of the surfaces (right); 
the isolated cavities then give way to a plate-spanning vapor tube (left), which can be larger than the critical vapor tube predicted by interfacial physics (dot-dashed line)~\cite{Remsing:PNAS:2015}.
Panels a and c adapted from Reference~\citenum{Vembanur:JPCB:2013} with permission; copyright 2013 American Chemical Society. 
Panel b adapted from reference~\citenum{Dhabal:JPCB:2021} with permission. 
Panel d adapted from Reference~\citenum{Jamadagni:JPCB:2009} with permission; copyright 2009 American Chemical Society. 
Panel e adapted from References~\citenum{Tiwary:PNAS:2015,Mondal:JSP:2020} with permission; copyright 2015 National Academy of Sciences. 
Panel g adapted from Reference~\citenum{Remsing:PNAS:2015} with permission; copyright 2015 National Academy of Sciences.
}
\label{fig:kinetics}
\end{figure}

Hydrophobic solutes, from noble gases and oils, to polymers and proteins, tend to adsorb to hydrophobic surfaces; 
localized at the surface, these solutes can nevertheless interact with one another and undergo assembly and disassembly. 
Vembanur et al.~\cite{Vembanur:JPCB:2013} showed that desolvation barriers, 
which impede the assembly of small non-polar solutes in bulk water, 
are significantly reduced near an extended hydrophobic surface
due to the relative ease with which its interfacial waters can be displaced (\fig{fig:kinetics}{c}).
Correspondingly, the kinetics of hydrophobic contact formation (and breakage) were found to be faster near hydrophobic surfaces, suggesting that such surfaces can generically catalyze the assembly (and disassembly) of small hydrophobic solutes.
Indeed, Jamadagni et al.~\cite{Jamadagni:JPCB:2009} showed that non-polar polymers, 
which tend to adopt globular conformations in bulk water,
instead form pancake-like structures with enhanced conformational fluctuations 
when adsorbed onto hydrophobic surfaces (\fig{fig:kinetics}{d}).
Extended hydrophobic surfaces have also been shown to facilitate protein unfolding and subsequent amyloid fibril formation~\cite{Pronchik:JACS:2010,Beverung:BC:1999}.
%

\FloatBarrier

\FloatBarrier

\subsection{Dewetting Barriers Impede the Interactions of Large Solutes}
The binding of either small or large hydrophobic solutes to extended hydrophobic surfaces 
is important in diverse contexts, ranging from protein interactions to nanoparticle self-assembly.
A number of studies have investigated the kinetics and mechanistic pathways of such processes, 
and highlighted that binding is hindered by dewetting barriers~\cite{Tiwary:PNAS:2015,Mondal:PNAS:2013,Setny:PNAS:2013,Tiwary:JCTC:2019}.
In particular, when two hydrophobic solutes approach one another, water confined between them becomes metastable with respect to its vapor~\cite{Berne:ARPC:2009}; 
the resulting bistability leads to slow solvent fluctuations between wet and dry states, 
and thereby to slow binding kinetics (\fig{fig:kinetics}{e}).
In studying the binding of methane to a hydrophobic pocket, Setny et al.~\cite{Setny:PNAS:2013} found that 
a substantial increase in a position-dependent friction coefficient (as methane approached the pocket) 
was needed to explain the slow binding kinetics.
Similarly, in their studies of fullerene binding to a hydrophobic pocket, Mondal et al.~\cite{Mondal:PNAS:2013}
found that considering both the ligand-pocket separation and water density in confinement provided a better description of the kinetics of hydrophobic assembly. 
Studying the same system, Tiwary et al.~\cite{Tiwary:PNAS:2015} found that relaxing steric constraints 
on the fullerene resulted in faster binding kinetics due to a lowering of the associated dewetting barriers.

\subsection{Dewetting in Hydrophobic Confinement}
Given its role in impeding assembly, the dewetting of the region between two hydrophobic surfaces,
fixed at a particular distance from one another, has been studied extensively~\cite{Huang:PNAS:2003,Choudhury:JACS:2007,Xu:JPCB:2010,Sharma:PNAS:2012,Altabet:PNAS:2016,Xi:MS:2018}.
According to classical interfacial physics, the distance between the hydrophobic solutes, $d_{\rm c}$, 
below which water confined between the surfaces becomes metastable with respect to its vapor (\fig{fig:kinetics}{f}), is proportional to solute size for nanoscopic solutes, and asymptotes to roughly $1.5~\mu$m for macroscopic solutes~\cite{Cerderina:JPCL:2011,Giovambattista:ARPC:2012}.
Moreover, the bottleneck to dewetting corresponds to the formation of a vapor tube, which spans the region between the surfaces~\cite{Lum:PRE:1997,Bolhuis:JCP:2000,Leung:PRL:2003} (\fig{fig:kinetics}{g}). 
Remsing et al.~\cite{Remsing:PNAS:2015} investigated the role of water density fluctuations in nucleating such vapor tubes, and found that enhanced fluctuations near hydrophobic surfaces stabilize isolated cavities adjacent to one or both confining surfaces, which then grow or coalesce to form vapor tubes;
moreover, the nascent vapor tubes can be larger than the critical vapor tubes predicted by interfacial physics.
Importantly, such a non-classical pathway, which circumvents the critical vapor tube, 
results in a lower barrier to dewetting.
Such non-classical dewetting pathways with smaller dewetting barriers have also been observed 
in the Wenzel-to-Cassie transition on rough hydrophobic surfaces~\cite{Quere:ARMR:2008,Papadopoulos:PNAS:2013,Arunachalam:JCIS:2019}, and have facilitated the design of textured surfaces, which can undergo barrierless dewetting 
and spontaneously recover their superhydrophobicity~\cite{Prakash:PNAS:2016}.

\section{FUTURE OUTLOOK}
The origins of our contemporary understanding of hydrophobic effects can be traced back to 
conceptual breakthroughs in the theory of liquids made more than 50 years ago by 
Widom, Stillinger, and others~\cite{Widom:Science:1967,Stillinger,Chandler:ARPC:2017}.
Indeed, many of the early insights into these fascinating phenomena were due to theoretical advances 
(e.g., scaled particle theory, Pratt-Chandler theory, information theory, Lum-Chandler-Weeks theory, etc.),
which focused on idealized non-polar solutes~\cite{Stillinger,Pratt:JCP:1977,Hummer:PNAS:1996,LCW}.
Concurrently, experiments focused on investigating manifestations of hydrophobic effects on more complex solutes, such as proteins, and were harder to interpret using theory~\cite{kauzmann,Snyder:PNAS:2011}.
The combination of recent theoretical, simulation, and experimental advances, 
described here, are beginning to bridge the divide between experiments and theory~\cite{Bellissent-Funel:CR:2016,Monroe:ARCBE:2020};
creative experiments are now being carried out on simpler, more idealized systems~\cite{Li:PNAS:2011,Davis:Nature:2012,Abbott:Nature:2015},
and molecular simulations are making it possible to investigate the implications of hydrophobic effects 
on more and more complex systems, from clay minerals and metals, to patterned surfaces and proteins~\cite{Rotenberg:JACS:2011,Limmer:PNAS:2013,Xi:PNAS:2017}.
These synergistic advances are poised to deliver exciting breakthroughs in biophysics, soft materials, and beyond.

%
Predicting the hydrophobicity of heterogeneous surfaces, using schemes that are both highly efficient and sufficiently accurate to predict binding affinities, has been a holy grail for both biophysics and soft materials design~\cite{Kister:PNAS:2008,Granick:Science:2008}.
The ability to accurately characterize the hydrophobicity of complex surfaces with nanoscale heterogeneity,
facilitated by an understanding of interfacial water density fluctuations (sec.~3), 
represents a necessary and important first step in this direction~\cite{Patel:PNAS:2011,Patel:JPCB:2012}.
Moreover, we now understand why well-intentioned prior efforts in this direction, such as the quest for an optimal hydropathy scale, were confounded --- the hydrophobicity on a heterogeneous surface depends sensitively on its curvature and chemical patterning (sec.~4)~\cite{Xi:PNAS:2017}.
In spite of the fact that the relationship between the hydrophobicity of a surface and its chemical and topographical features is both incredibly complex and high dimensional, recent advances in data science may nevertheless provide the requisite framework for capturing such a relationship~\cite{Monroe:PNAS:2018,vanlehn:JPCB:2020}.
In particular, a sufficiently large and appropriately curated library of diverse heterogeneous surfaces and/or proteins, whose context-dependent hydrophobicity has been characterized using enhanced sampling techniques or high-throughput experiments, could be used to train a supervised machine learning model.

%
Learning the complex relationship between the hydrophobicity of a surface and its chemical and topographical patterns could facilitate the design of nano-structured surfaces with bespoke hydrophobicity.
For example, the identification of chemical patterns, which confer heterogeneous surfaces with super-hydrophilicity, i.e., exceedingly strong interactions with water, could facilitate the discovery of surfaces that resist fouling and are superoleophobic underwater~\cite{Chen:Biomaterials:2009}.
The ability to characterize the hydrophobicity of proteins and an appreciation of its role in 
driving protein interactions and assemblies (sec.~5)~\cite{Rego:PNAS:2021} has also set the stage for 
identifying aggregation-prone proteins regions and engineering proteins with enhanced solubility~\cite{Shire:JPS:2004,Chong:AC:2014}. 
Moreover, an understanding of the role water density fluctuations in hindering assembly (sec.~6),
e.g., in protein folding or supramolecular host-guest interactions,
could pave the way for engineering assembly pathways and exercising control over the kinetics of assembly~\cite{Jiang:JPCB:2019,Dhabal:JPCB:2021,Tiwary:PNAS:2015}.
Finally, an understanding of dewetting pathways in hydrophobic confinement could also facilitate the rational design of textured hydrophobic materials that are capable of dewetting spontaneously, even under hydrostatic pressure, thereby paving the way for their use underwater and in condensation heat transfer~\cite{Prakash:PNAS:2016}.


\section*{DISCLOSURE STATEMENT}
The authors are not aware of any affiliations, memberships, funding, or financial holdings that
might be perceived as affecting the objectivity of this review.

\section*{ACKNOWLEDGMENTS}
%
A.J.P. gratefully acknowledges financial support from the National Science Foundation (CBET 1652646, CHE 1665339, and UPENN MRSEC DMR 1720530), and awards from the Alfred P. Sloan Research Foundation (FG-2017-9406) and the Camille and Henry Dreyfus Foundation (TC-19-033). 
N.B.R. was supported by the National Science Foundation grants DMR 1844514 and CHE 1665339.
The authors thank Jagannath Mondal, Richard Remsing and Pratyush Tiwary for their feedback on an early version of the manuscript.
A.J.P. is grateful for numerous insightful discussions with Shekhar Garde and Hank Ashbaugh.
%

\bibliographystyle{ar-style4}


\end{document}